# Efficient simulation for light scattering from plasmonic core-shell nanospheres on a substrate for biosensing

Huai-Yi Xie,[1,2] Minfeng Chen,[1] Yia-Chung Chang,[1,3,*] and Rakesh Singh Moirangthem[1,4]

[1]*Research Center for Applied Sciences, Academia Sinica, 128 Academia Road, Sec. 2, Taipei, 1529 Taiwan*

[2]*National Center for Theoretical Sciences (South), National Cheng Kung University, Tainan, 701 Taiwan*

[3]*Department of Physics, National Cheng Kung University, Tainan, 701 Taiwan*

[4]*Department of Applied Physics, Indian School of Mines, Dhanbad - 826004, Jharkhand, India*

**Abstract**

We have developed an efficient numerical method to investigate light scattering from plasmonic nanospheres on a substrate covered by a shell, based on the half-space Green's function approach. We use this method to study optical scattering from DNA molecules attached to metallic nanoparticles on a substrate and compare with experiment. We obtain fairly good agreement between theoretical predictions and the measured ellipsometric spectra. The metallic nanoparticles were used to detect the binding with DNA molecules in a microfluidic setup via spectroscopic ellipsometry (SE), and a detectable change in ellipsometric spectra was found when DNA molecules are captured on Au nanoparticles surface. Our theoretical simulation indicates that the coverage of Au nanosphere by a submonolayer of DNA molecules, which is modeled by a thin layer of dielectric material, can indeed lead to a small but detectable change in ellipsometric spectra. Our studies demonstrated the ultra-sensitive capability of SE for sensing submonolayer coverage of DNA molecules on Au nanospheres.

Corresponding author: *yiachang@gate.sinica.edu.tw

# I. Introduction

In recent years, spectroscopic ellipsometry (SE) has been wildly used as an optical metrology tool in IC industry [1] and for biosensing [2-6]. In biosensing, detecting the DNA sequences plays an important role in revealing the genetic codes. One popular detecting method is to attach various DNA sequences on Au nanoparticles (AuNPs). [7-14] Many applications of the surface plasmonic effect for metallic nanoparticles have been discussed in the literature.[15-19] Due to the effect of localized surface plasmon resonance (SPR), the detection signal can be amplified significantly so that one can learn some information of the DNA sequences. [20]

Theoretically, efficient simulation software has been developed, which can be used to determine the critical dimension of nanostructures under inspection. Rather than using approximate method such as the effective medium approach, [21-23] we adopt the volume integral method based on the half-space Green's function. [24] The advantage of this method is that we only have to solve the electromagnetic field in the integration domain and then use the Lippman-Schwinger equation to obtain the full solution everywhere. This can save a lot of computation resources and time compared with other numerical methods such as finite-difference time-domain (FDTD) and finite-element method (FEM) methods when the integration domain is small. Furthermore, we improve the efficiency and accuracy of this method by choosing geometry-adopted basis functions rather than polynomial basis as adopted in [24]. This is of particular importance for frequencies near the plasmonic resonance of metallic nanoparticles where a large number of basis functions in the polynomial form is required to obtain reliable results. [25] For example, if the shape of nanoparticles is spherical, we choose the products of spherical harmonics and spherical Bessel functions as the basis functions. [26,27] This significantly reduces the number of basis functions needed to describe the localized nature of the electromagnetic field for frequencies near the plasmonic resonance. [26] Using this geometry-adopted basis, we have analyzed the effect of clustering features on the ellipsometry spectra of glass substrates covered with a random distribution of Au nanoparticles (AuNPs) and determined the structure information of the sample, including average sizes, separations of nanoparticles and fractional area covered by clusters of AuNPs. [27]

In this paper, we first construct the theoretical formulation for light scattering from an isolated nanoparticle first fully and next partially covered by a shell, using Green's function (GF) method with spherical harmonic basis functions. Next we compare our theoretical predictions for an isolated nanoparticle fully covered by a shell with the conventional core-shell Mie scattering theory in the absence of a substrate. Finally, we apply our theoretical method to the sensing of DNA molecules attached to Au nanoparticles by calculating the ellipsometry spectra (i.e. $\Psi$ and $\Delta$) and compare with corresponding experimental data. The comparison between theory and experiment allows us to find a suitable description of the critical dimension and average optical constant of biological objects sensed by the Au nanoparticles placed on a BK7 prism. Preliminary experimental results for ellipsometry spectra of DNA covered Au nanoparticles were reported in [28], but no suitable theoretical interpretation was available. Here, we apply the newly developed theoretical analysis based on the half-space Green's function method and make comparison with the experimental data in order to confirm the capability of SE sensing for DNA molecules covering Au nanoparticles. We model the DNA coverage by a thin dielectric shell which encloses the metallic nanosphere. Our theoretical method is compared with the finite-element method (FEM) used in the popular commercial package COMSOL for a periodic array of core-shell particles. Furthermore, we also simulated the ellipsometric spectra for a random distribution of core-shell nanoparticles and found better agreement with experimental data. In our modeling, the input dielectric constants of both Au and Ag are taken from experimental values provided in Ref. [29].

## II. Theoretical formulation of light scattering from an isolated/periodic/random core-shell nanoparticle on a substrate

We start by writing the Lippmann-Schwinger (L-S) equation for the electric field of an isolated core-shell nanosphere on a substrate (shown in Fig. 1) as the following form:

$$\mathbf{E}(\mathbf{r}) = \mathbf{E}_0(\mathbf{r}) + k_0^2(\varepsilon_1 - \varepsilon_a)\int_{\Omega_1} \mathbf{G}(\mathbf{r},\mathbf{r}')\cdot\mathbf{E}(\mathbf{r}')d^3\mathbf{r}' + k_0^2(\varepsilon_2 - \varepsilon_a)\int_{\Omega_2} \mathbf{G}(\mathbf{r},\mathbf{r}')\cdot\mathbf{E}(\mathbf{r}')d^3\mathbf{r}',$$

(1)

where $E_{0j}(\mathbf{r}) = T_{0j}e^{i\mathbf{k}_0\cdot\mathbf{r}} + R_{0j}e^{i\mathbf{k}_0'\cdot\mathbf{r}}$ is the electric field of the non-perturbed system (i.e. multilayer system in the absence of core-shell nanoparticles) with $\mathbf{k}_0$ ($\mathbf{k}_0'$) denoting an incident (reflected) wave vector in the ambient $[(\mathbf{k}_0')_z = -(\mathbf{k}_0)_z]$. $R_{0j}$ and $T_{0j}$ denote the $j$-component ($j = x, y, z$) reflection and transmission coefficients for the multilayer system without scatterers. $\varepsilon_i$ is the dielectric function in region $\Omega_i$; $i = 1,2$ labels the inner and outer regions, and $R_1$ and $R_2$ are the inner and outer radii of the core-shell. $\varepsilon_a$, $\varepsilon_g$ and $\varepsilon_{sub}$ denote the dielectric constants of the ambient, the host material of the grating layer (which contains scatterers), and the substrate, respectively. For isolated distribution, the form of dyadic Green's function is $\mathbf{G}(\mathbf{r},\mathbf{r}') = \frac{1}{(2\pi)^2} \int d\mathbf{k}_n e^{i\mathbf{k}_n \cdot (\boldsymbol{\rho}-\boldsymbol{\rho}')} \mathbf{g}_n(z,z')$ which has been shown in Eq. (3) of Ref. [26]. Note that the vector $\mathbf{k}_n$ denotes arbitrary wave-vector in the x-y plane.[26] For periodic distribution, we have $\mathbf{G}(\mathbf{r},\mathbf{r}') = \sum_n e^{i\mathbf{k}_n \cdot (\boldsymbol{\rho}-\boldsymbol{\rho}')} \mathbf{g}_n(z,z')$ as shown in Eq. (4) of Ref. [24] with $\mathbf{k}_n = \mathbf{k}_0 + \mathbf{g}_n$, where $n$ is a finite number, $\mathbf{k}_0$ is the incident wave-vector, and $\mathbf{g}_n$ is the reciprocal lattice vectors. [24] Furthermore, we make the expansion

$$\mathbf{E}(\mathbf{r}) = \begin{cases} \sum_{\ell m} \boldsymbol{\alpha}_{\ell m} j_\ell(k_1 r) Y_{\ell m}(\Omega), \mathbf{r} \in \Omega_1 \\ \sum_{\ell m} \left[ \boldsymbol{\beta}_{\ell m} j_\ell(k_2 r) + \boldsymbol{\gamma}_{\ell m} n_\ell(k_2 r) \right] Y_{\ell m}(\Omega), \mathbf{r} \in \Omega_2 \end{cases}, \quad (2)$$

where $j_\ell$ is the spherical Bessel function of order $\ell$, $n_\ell$ is the spherical Nummen function of order $\ell$ and $Y_{\ell m}$ is the spherical harmonics with quantum numbers $(\ell,m)$. $k_i$ is the wavenumber in region $\Omega_i, i=1,2$. Substituting Eq. (2) into the integral equation and projecting Eq. (1) into the basis functions $j_\ell(k_1 r) Y_{\ell m}(\Omega)$ in the region $\Omega_1$ and $j_\ell(k_2 r) Y_{\ell m}(\Omega)$, $n_\ell(k_2 r) Y_{\ell m}(\Omega)$ in region $\Omega_2$ respectively, we obtain a set of coupled equations for solving the unknown coefficients $\boldsymbol{\alpha}_{\ell m}, \boldsymbol{\beta}_{\ell m}$, and $\boldsymbol{\gamma}_{\ell m}$:

$$\begin{cases} \sum_{\ell'm'} I^{11}_{\ell m,\ell'm'} \alpha^j_{\ell'm'} = F^{1\ell m}_j + \\ (\varepsilon_1 - \varepsilon_a) \sum_{j'\ell'm'} \mathbf{G}^{11,\ell m \ell'm'}_{jj'} \alpha^{j'}_{\ell'm'} + (\varepsilon_2 - \varepsilon_a) \sum_{j'\ell'm'} \mathbf{G}^{12,\ell m \ell'm'}_{jj'} \beta^{j'}_{\ell'm'} + (\varepsilon_2 - \varepsilon_a) \sum_{j'\ell'm'} \mathbf{G}^{13,\ell m \ell'm'}_{jj'} \gamma^{j'}_{\ell'm'} \\ \sum_{\ell'm'} I^{22}_{\ell m,\ell'm'} \beta^j_{\ell'm'} + \sum_{\ell'm'} I^{23}_{\ell m,\ell'm'} \gamma^j_{\ell'm'} = F^{2\ell m}_j + \\ (\varepsilon_1 - \varepsilon_a) \sum_{j'\ell'm'} \mathbf{G}^{21,\ell m \ell'm'}_{jj'} \alpha^{j'}_{\ell'm'} + (\varepsilon_2 - \varepsilon_a) \sum_{j'\ell'm'} \mathbf{G}^{22,\ell m \ell'm'}_{jj'} \beta^{j'}_{\ell'm'} + (\varepsilon_2 - \varepsilon_a) \sum_{j'\ell'm'} \mathbf{G}^{23,\ell m \ell'm'}_{jj'} \gamma^{j'}_{\ell'm'} \\ \sum_{\ell'm'} I^{32}_{\ell m,\ell'm'} \beta^j_{\ell'm'} + \sum_{\ell'm'} I^{33}_{\ell m,\ell'm'} \gamma^j_{\ell'm'} I^{33}_\ell = F^{3\ell m}_j + \\ (\varepsilon_1 - \varepsilon_a) \sum_{j'\ell'm'} \mathbf{G}^{31,\ell m \ell'm'}_{jj'} \alpha^{j'}_{\ell'm'} + (\varepsilon_2 - \varepsilon_a) \sum_{j'\ell'm'} \mathbf{G}^{32,\ell m \ell'm'}_{jj'} \beta^{j'}_{\ell'm'} + (\varepsilon_2 - \varepsilon_a) \sum_{j'\ell'm'} \mathbf{G}^{33,\ell m \ell'm'}_{jj'} \gamma^{j'}_{\ell'm'} \end{cases}$$

(3)

where for the case as shown in Fig. 1(a), we have the following forms:

$$I^{11}_{\ell m,\ell'm'} = \delta_{\ell\ell'} \delta_{mm'} \int_0^{R_1} dr r^2 \left[ j_\ell(k_1 r) \right]^* j_\ell(k_1 r), \quad (4a)$$

$$I^{22}_{\ell m,\ell'm'} = \delta_{\ell\ell'}\delta_{mm'}\int_{R_1}^{R_2} dr\, r^2 \left[j_\ell(k_2 r)\right]^* j_\ell(k_2 r), \tag{5a}$$

$$I^{23}_{\ell m,\ell'm'} = \delta_{\ell\ell'}\delta_{mm'}\int_{R_1}^{R_2} dr\, r^2 \left[j_\ell(k_2 r)\right]^* n_\ell(k_2 r), \tag{6a}$$

$$I^{32}_{\ell m,\ell'm'} = \delta_{\ell\ell'}\delta_{mm'}\int_{R_1}^{R_2} dr\, r^2 \left[n_\ell(k_2 r)\right]^* j_\ell(k_2 r), \tag{7a}$$

$$I^{33}_{\ell m,\ell'm'} = \delta_{\ell\ell'}\delta_{mm'}\int_{R_1}^{R_2} dr\, r^2 \left[n_\ell(k_2 r)\right]^* n_\ell(k_2 r). \tag{8a}$$

The superscript * in the above equations denotes taking the complex conjugate.

Next for the case as shown in Fig. 1(b), we have the following forms:

$$I^{11}_{\ell m,\ell'm'} =$$

$$\int_{-R_1}^{R_1} dz \int_0^{\sqrt{R_1^2-z^2}} \rho\, d\rho \left[j_\ell\left(k_1\sqrt{\rho^2+z^2}\right)\right]^* \tilde{P}_{\ell m}\left(\frac{z}{\sqrt{\rho^2+z^2}}\right) j_{\ell'}\left(k_1\sqrt{\rho^2+z^2}\right) \tilde{P}_{\ell'm}\left(\frac{z}{\sqrt{\rho^2+z^2}}\right) \delta_{mm'},$$
$$\tag{4b}$$

$$I^{22}_{\ell m,\ell'm'} =$$

$$\int_{-R_1}^{R_1} dz \int_{\sqrt{R_1^2-z^2}}^{\sqrt{R_2^2-z^2}} \rho\, d\rho \left[j_\ell\left(k_2\sqrt{\rho^2+z^2}\right)\right]^* \tilde{P}_{\ell m}\left(\frac{z}{\sqrt{\rho^2+z^2}}\right) j_{\ell'}\left(k_2\sqrt{\rho^2+z^2}\right) \tilde{P}_{\ell'm}\left(\frac{z}{\sqrt{\rho^2+z^2}}\right) \delta_{mm'}$$

$$+\int_{R_1}^{R_2} dz \int_0^{\sqrt{R_2^2-z^2}} \rho\, d\rho \left[j_\ell\left(k_2\sqrt{\rho^2+z^2}\right)\right]^* \tilde{P}_{\ell m}\left(\frac{z}{\sqrt{\rho^2+z^2}}\right) j_{\ell'}\left(k_2\sqrt{\rho^2+z^2}\right) \tilde{P}_{\ell'm}\left(\frac{z}{\sqrt{\rho^2+z^2}}\right) \delta_{mm'},$$
$$\tag{5b}$$

$$I^{23}_{\ell m,\ell'm'} =$$

$$\int_{-R_1}^{R_1} dz \int_{\sqrt{R_1^2-z^2}}^{\sqrt{R_2^2-z^2}} \rho\, d\rho \left[j_\ell\left(k_2\sqrt{\rho^2+z^2}\right)\right]^* \tilde{P}_{\ell m}\left(\frac{z}{\sqrt{\rho^2+z^2}}\right) n_{\ell'}\left(k_2\sqrt{\rho^2+z^2}\right) \tilde{P}_{\ell'm}\left(\frac{z}{\sqrt{\rho^2+z^2}}\right) \delta_{mm'}$$

$$+\int_{R_1}^{R_2} dz \int_0^{\sqrt{R_2^2-z^2}} \rho\, d\rho \left[j_\ell\left(k_2\sqrt{\rho^2+z^2}\right)\right]^* \tilde{P}_{\ell m}\left(\frac{z}{\sqrt{\rho^2+z^2}}\right) n_{\ell'}\left(k_2\sqrt{\rho^2+z^2}\right) \tilde{P}_{\ell'm}\left(\frac{z}{\sqrt{\rho^2+z^2}}\right) \delta_{mm'},$$
$$\tag{6b}$$

$$I^{32}_{\ell m,\ell'm'} =$$

$$\int_{-R_1}^{R_1} dz \int_{\sqrt{R_1^2-z^2}}^{\sqrt{R_2^2-z^2}} \rho\, d\rho \left[n_\ell\left(k_2\sqrt{\rho^2+z^2}\right)\right]^* \tilde{P}_{\ell m}\left(\frac{z}{\sqrt{\rho^2+z^2}}\right) j_{\ell'}\left(k_2\sqrt{\rho^2+z^2}\right) \tilde{P}_{\ell'm}\left(\frac{z}{\sqrt{\rho^2+z^2}}\right) \delta_{mm'}$$

$$+\int_{R_1}^{R_2} dz \int_0^{\sqrt{R_2^2-z^2}} \rho\, d\rho \left[n_\ell\left(k_2\sqrt{\rho^2+z^2}\right)\right]^* \tilde{P}_{\ell m}\left(\frac{z}{\sqrt{\rho^2+z^2}}\right) j_{\ell'}\left(k_2\sqrt{\rho^2+z^2}\right) \tilde{P}_{\ell'm}\left(\frac{z}{\sqrt{\rho^2+z^2}}\right) \delta_{mm'}$$

, (7b)

$$I^{33}_{\ell m, \ell'm'} =$$

$$\int_{-R_1}^{R_1} dz \int_{\sqrt{R_1^2-z^2}}^{\sqrt{R_2^2-z^2}} \rho d\rho \left[ n_\ell \left( k_2 \sqrt{\rho^2+z^2} \right) \right]^* \tilde{P}_{\ell m} \left( \frac{z}{\sqrt{\rho^2+z^2}} \right) n_{\ell'} \left( k_2 \sqrt{\rho^2+z^2} \right) \tilde{P}_{\ell'm} \left( \frac{z}{\sqrt{\rho^2+z^2}} \right) \delta_{mm'}$$

$$+ \int_{R_1}^{R_2} dz \int_0^{\sqrt{R_2^2-z^2}} \rho d\rho \left[ n_\ell \left( k_2 \sqrt{\rho^2+z^2} \right) \right]^* \tilde{P}_{\ell m} \left( \frac{z}{\sqrt{\rho^2+z^2}} \right) n_{\ell'} \left( k_2 \sqrt{\rho^2+z^2} \right) \tilde{P}_{\ell'm} \left( \frac{z}{\sqrt{\rho^2+z^2}} \right) \delta_{mm'}$$

, (8b)

and $\mathbf{G}^{pq,\ell m\ell'm'}_{j\gamma} = \sum_{n\alpha\beta} \mathbf{S}_{n,\alpha j} \bar{\mathbf{G}}^{pq,\ell m\ell'm'}_{n,\alpha\beta} \mathbf{S}_{n,\beta\gamma}$, $p,q = 1,2,3$ and $S$ is a rotation matrix as defined in Eq. (5) of Ref. [27]. Matrix elements $\bar{\mathbf{G}}^{pq,\ell m\ell'm'}_{n,\alpha\beta}$ for the case of a sphere have been given in Ref. [27]. Here, we extend to the core-shell case. The three components of $F_j^{p\ell m}$, $p = 1,2,3$ are:

For the case as shown in Fig. 1(a):

$$F_j^{1\ell m} = 4\pi i^\ell \int_0^{R_1} dr r^2 \left[ j_\ell(k_1 r) \right]^* j_\ell(k_0 r) \left[ T_{0j} e^{ik_{0z}R_2} Y^*_{\ell m}(\Omega_{k_0}) + R_{0j} e^{-ik_{0z}R_2} Y^*_{\ell m}(\pi - \theta_{k_0}, \varphi_{k_0}) \right],$$

(9a)

$$F_j^{2\ell m} = 4\pi i^\ell \int_{R_1}^{R_2} dr r^2 \left[ j_\ell(k_2 r) \right]^* j_\ell(k_0 r) \left[ T_{0j} e^{ik_{0z}R_2} Y^*_{\ell m}(\Omega_{k_0}) + R_{0j} e^{-ik_{0z}R_2} Y^*_{\ell m}(\pi - \theta_{k_0}, \varphi_{k_0}) \right],$$

(10a)

$$F_j^{3\ell m} = 4\pi i^\ell \int_{R_1}^{R_2} dr r^2 \left[ n_\ell(k_2 r) \right]^* j_\ell(k_0 r) \left[ T_{0j} e^{ik_{0z}R_2} Y^*_{\ell m}(\Omega_{k_0}) + R_{0j} e^{-ik_{0z}R_2} Y^*_{\ell m}(\pi - \theta_{k_0}, \varphi_{k_0}) \right],$$

(11a)

For the case as shown in Fig. 1(b):

$$F_j^{1\ell m} = \left[ T_{0j} e^{ik_{0z}R_1} + (-1)^{\ell+m} R_{0j} e^{-ik_{0z}R_1} \right] \sqrt{2\pi} (-1)^m e^{-im\frac{\pi}{2}} \times$$

$$\int_{-R_1}^{R_1} dz e^{ik_{0z}z} \int_0^{\sqrt{R_1^2-z^2}} \rho d\rho \left[ j_\ell \left( k_1 \sqrt{\rho^2+z^2} \right) \right]^* \tilde{P}_{\ell m} \left( \frac{z}{\sqrt{\rho^2+z^2}} \right) J_m(k_{0x}\rho),$$

(9b)

$$F_j^{2\ell m} = \left[T_{0j}e^{ik_{0z}R_1} + (-1)^{\ell+m} R_{0j}e^{-ik_{0z}R_1}\right]\sqrt{2\pi}(-1)^m e^{-im\frac{\pi}{2}} \times$$

$$\int_{-R_1}^{R_1} dz e^{ik_{0z}z} \int_{\sqrt{R_1^2-z^2}}^{\sqrt{R_2^2-z^2}} \rho d\rho \left[j_\ell\left(k_2\sqrt{\rho^2+z^2}\right)\right]^* \tilde{P}_{\ell m}\left(\frac{z}{\sqrt{\rho^2+z^2}}\right) J_m(k_{0x}\rho)$$

$$+\left[T_{0j}e^{ik_{0z}R_1} + (-1)^{\ell+m} R_{0j}e^{-ik_{0z}R_1}\right]\sqrt{2\pi}(-1)^m e^{-im\frac{\pi}{2}} \times$$

$$\int_{R_1}^{R_2} dz e^{ik_{0z}z} \int_0^{\sqrt{R_2^2-z^2}} \rho d\rho \left[j_\ell\left(k_2\sqrt{\rho^2+z^2}\right)\right]^* \tilde{P}_{\ell m}\left(\frac{z}{\sqrt{\rho^2+z^2}}\right) J_m(k_{0x}\rho)$$

(10b)

$$F_j^{3\ell m} = \left[T_{0j}e^{ik_{0z}R_1} + (-1)^{\ell+m} R_{0j}e^{-ik_{0z}R_1}\right]\sqrt{2\pi}(-1)^m e^{-im\frac{\pi}{2}} \times$$

$$\int_{-R_1}^{R_1} dz e^{ik_{0z}z} \int_{\sqrt{R_1^2-z^2}}^{\sqrt{R_2^2-z^2}} \rho d\rho \left[n_\ell\left(k_2\sqrt{\rho^2+z^2}\right)\right]^* \tilde{P}_{\ell m}\left(\frac{z}{\sqrt{\rho^2+z^2}}\right) J_m(k_{0x}\rho)$$

$$+\left[T_{0j}e^{ik_{0z}R_1} + (-1)^{\ell+m} R_{0j}e^{-ik_{0z}R_1}\right]\sqrt{2\pi}(-1)^m e^{-im\frac{\pi}{2}} \times$$

$$\int_{R_1}^{R_2} dz e^{ik_{0z}z} \int_0^{\sqrt{R_2^2-z^2}} \rho d\rho \left[n_\ell\left(k_2\sqrt{\rho^2+z^2}\right)\right]^* \tilde{P}_{\ell m}\left(\frac{z}{\sqrt{\rho^2+z^2}}\right) J_m(k_{0x}\rho)$$

(11b)

where $k_{0x}$ ($k_{0z}$) is the x(z)-component of wave number $\mathbf{k}_0$ outside the core-shell sphere (air for the present case) and $q_n \equiv \sqrt{k_n^2-k_0^2}$. $R_n$ (or $\tilde{R}_n$) denotes the reflection matrix defined in [24], including the factor $e^{-4q_n R_2}$ for Fig. 1(a) and $e^{-4q_n R_1}$ for Fig. 1(b), since we have chosen z axis to be pointing downward and z=0 at the center of the nanosphere. $R_{0j}$ and $T_{0j}, j=x,y,z$ denote the j-component of the reflective and transmission coefficients in the grating layer in the absence of core-shell sphere. The following integrals are needed for describing the matrix elements $\bar{\mathbf{G}}_{n,\alpha\beta}^{pq,\ell m\ell'm'}$ (see Eqs. (6)-(10) of Ref. [27]):

For the case as shown in Fig. 1(a):

$$E_{n\ell m}^{1\pm} = e^{im\tilde{\varphi}_n} \int_{-R_1}^{R_1} dz e^{\pm q_n z} e^{-q_n R_2} \left[I_{n\ell m}^1(z)\right]^*$$

$$E_{n\ell m}^{2\pm} = e^{im\tilde{\varphi}_n} \left[\int_{-R_2}^{-R_1} dz e^{\pm q_n z} e^{-q_n R_2} \left[I_{n\ell m}^{2a}(z)\right]^* + \int_{-R_1}^{R_1} dz e^{\pm q_n z} e^{-q_n R_2} \left[I_{n\ell m}^{2b}(z)\right]^* + \int_{R_1}^{R_2} dz e^{\pm q_n z} e^{-q_n R_2} \left[I_{n\ell m}^{2a}(z)\right]^*\right]$$

$$E_{n\ell m}^{3\pm} = e^{im\tilde{\varphi}_n} \left[\int_{-R_2}^{-R_1} dz e^{\pm q_n z} e^{-q_n R_2} \left[I_{n\ell m}^{3a}(z)\right]^* + \int_{-R_1}^{R_1} dz e^{\pm q_n z} e^{-q_n R_2} \left[I_{n\ell m}^{3b}(z)\right]^* + \int_{R_1}^{R_2} dz e^{\pm q_n z} e^{-q_n R_2} \left[I_{n\ell m}^{3a}(z)\right]^*\right]$$

, (12a)

$$B^{1\pm}_{n\ell'm'} = e^{-im'\tilde{\varphi}_n} \int_{-R_1}^{R_1} dz' e^{\pm q_n z'} e^{-q_n R_2} I^{1}_{n\ell'm'}(z')$$

$$B^{2\pm}_{n\ell'm'} = e^{-im'\tilde{\varphi}_n} \left[ \int_{-R_2}^{-R_1} dz' e^{\pm q_n z'} e^{-q_n R_2} I^{2a}_{n\ell'm'}(z') + \int_{-R_1}^{R_1} dz' e^{\pm q_n z'} e^{-q_n R_2} I^{2b}_{n\ell'm'}(z') + \int_{R_1}^{R_2} dz' e^{\pm q_n z'} e^{-q_n R_2} I^{2a}_{n\ell'm'}(z') \right]$$

$$B^{3\pm}_{n\ell'm'} = e^{-im'\tilde{\varphi}_n} \left[ \int_{-R_2}^{-R_1} dz' e^{\pm q_n z'} e^{-q_n R_2} I^{3a}_{n\ell'm'}(z') + \int_{-R_1}^{R_1} dz' e^{\pm q_n z'} e^{-q_n R_2} I^{3b}_{n\ell'm'}(z') + \int_{R_1}^{R_2} dz' e^{\pm q_n z'} e^{-q_n R_2} I^{3a}_{n\ell'm'}(z') \right]$$

(13a)

For the case as shown in Fig. 1(b):

$$E^{1\pm}_{n\ell m} = e^{im\tilde{\varphi}_n} \int_{-R_1}^{R_1} dz e^{\pm q_n z} e^{-q_n R_1} \left[ I^{1}_{n\ell m}(z) \right]^*$$

$$E^{2\pm}_{n\ell m} = e^{im\tilde{\varphi}_n} \left[ \int_{-R_1}^{R_1} dz e^{\pm q_n z} e^{-q_n R_1} \left[ I^{2b}_{n\ell m}(z) \right]^* + \int_{R_1}^{R_2} dz e^{\pm q_n z} e^{-q_n R_1} \left[ I^{2a}_{n\ell m}(z) \right]^* \right],$$
(12b)

$$E^{3\pm}_{n\ell m} = e^{im\tilde{\varphi}_n} \left[ \int_{-R_1}^{R_1} dz e^{\pm q_n z} e^{-q_n R_1} \left[ I^{3b}_{n\ell m}(z) \right]^* + \int_{R_1}^{R_2} dz e^{\pm q_n z} e^{-q_n R_1} \left[ I^{3a}_{n\ell m}(z) \right]^* \right]$$

$$B^{1\pm}_{n\ell'm'} = e^{-im'\tilde{\varphi}_n} \int_{-R_1}^{R_1} dz' e^{\pm q_n z'} e^{-q_n R_1} I^{1}_{n\ell'm'}(z')$$

$$B^{2\pm}_{n\ell'm'} = e^{-im'\tilde{\varphi}_n} \left[ \int_{R_1}^{R_2} dz' e^{\pm q_n z'} e^{-q_n R_1} I^{2a}_{n\ell'm'}(z') + \int_{-R_1}^{R_1} dz' e^{\pm q_n z'} e^{-q_n R_1} I^{2b}_{n\ell'm'}(z') \right],$$
(13b)

$$B^{3\pm}_{n\ell'm'} = e^{-im'\tilde{\varphi}_n} \left[ \int_{R_1}^{R_2} dz' e^{\pm q_n z'} e^{-q_n R_1} I^{3a}_{n\ell'm'}(z') + \int_{-R_1}^{R_1} dz' e^{\pm q_n z'} e^{-q_n R_1} I^{3b}_{n\ell'm'}(z') \right]$$

where $\tilde{\varphi}_n = \tan^{-1}\dfrac{k_{xn}}{k_{yn}}$ and

$$I^{1}_{n\ell m}(z) = \sqrt{2\pi} \int_{0}^{\sqrt{R_1^2 - z^2}} d\rho \rho j_\ell\left(k_1\sqrt{\rho^2 + z^2}\right) \tilde{P}_{\ell m}\left(\frac{z}{\sqrt{\rho^2 + z^2}}\right) J_m(k_n \rho). \tag{14}$$

$$\begin{cases} I^{2a}_{n\ell m}(z) = \sqrt{2\pi} \int_{0}^{\sqrt{R_2^2 - z^2}} d\rho \rho j_\ell\left(k_2\sqrt{\rho^2 + z^2}\right) \tilde{P}_{\ell m}\left(\dfrac{z}{\sqrt{\rho^2 + z^2}}\right) J_m(k_n \rho) \\ I^{3a}_{n\ell m}(z) = \sqrt{2\pi} \int_{0}^{\sqrt{R_2^2 - z^2}} d\rho \rho n_\ell\left(k_2\sqrt{\rho^2 + z^2}\right) \tilde{P}_{\ell m}\left(\dfrac{z}{\sqrt{\rho^2 + z^2}}\right) J_m(k_n \rho) \end{cases}. \tag{15}$$

$$\begin{cases} I^{2b}_{n\ell m}(z) = \sqrt{2\pi} \int_{\sqrt{R_1^2 - z^2}}^{\sqrt{R_2^2 - z^2}} d\rho \rho j_\ell\left(k_2\sqrt{\rho^2 + z^2}\right) \tilde{P}_{\ell m}\left(\dfrac{z}{\sqrt{\rho^2 + z^2}}\right) J_m(k_n \rho) \\ I^{3b}_{n\ell m}(z) = \sqrt{2\pi} \int_{\sqrt{R_1^2 - z^2}}^{\sqrt{R_2^2 - z^2}} d\rho \rho n_\ell\left(k_2\sqrt{\rho^2 + z^2}\right) \tilde{P}_{\ell m}\left(\dfrac{z}{\sqrt{\rho^2 + z^2}}\right) J_m(k_n \rho) \end{cases}. \tag{16}$$

where $\tilde{P}_{\ell m}$ is the normalized associated Legendre function with quantum number $(\ell, m)$, and $J_m$ is the Bessel function of order $m$. For the singular terms involving $e^{-q_n|z-z'|}$ and $\text{sgn}(z-z')e^{-q_n|z-z'|}$, we also need to evaluate the integrals $M^{pq+}_{n\ell m \ell'm'}$ and

$M^{pq-}_{n\ell m\ell'm'}$, $p,q = 1,2,3$. We have:

For the case as shown in Fig. 1(a):

$$M^{11\pm}_{n\ell m\ell'm'} = e^{im\tilde{\varphi}_n}e^{-im'\tilde{\varphi}_n}\int_{-R_1}^{R_1}dz\left[I^1_{n\ell m}(z)\right]^*\left[\int_{-R_1}^{z}dz'e^{-q_n(z-z')}I^1_{n\ell'm'}(z')\pm\int_{z}^{R_1}dz'e^{-q_n(z'-z)}I^1_{n\ell'm'}(z')\right],$$

(17a)

$$M^{1q\pm}_{n\ell m\ell'm'} = e^{im\tilde{\varphi}_n}e^{-im'\tilde{\varphi}_n}\int_{-R_1}^{R_1}dz\left[I^1_{n\ell m}(z)\right]^*\left[\int_{-R_1}^{z}dz'e^{-q_n(z-z')}I^{qb}_{n\ell'm'}(z')\pm\int_{z}^{R_1}dz'e^{-q_n(z'-z)}I^{qb}_{n\ell'm'}(z')\right]$$
$$+e^{im\tilde{\varphi}_n}e^{-im'\tilde{\varphi}_n}\int_{-R_1}^{R_1}dz\left[I^1_{n\ell m}(z)\right]^*e^{-q_nz}\int_{-R_2}^{-R_1}dz'e^{q_nz'}I^{qa}_{n\ell'm'}(z')$$
$$\pm e^{im\tilde{\varphi}_n}e^{-im'\tilde{\varphi}_n}\int_{-R_1}^{R_1}dz\left[I^1_{n\ell m}(z)\right]^*e^{q_nz}\int_{R_1}^{R_2}dz'e^{-q_nz'}I^{qa}_{n\ell'm'}(z'); q=2,3$$

(18a)

$$M^{p1\pm}_{n\ell m\ell'm'} = e^{im\tilde{\varphi}_n}e^{-im'\tilde{\varphi}_n}\int_{-R_1}^{R_1}dz\left[I^{pb}_{n\ell m}(z)\right]^*\left[\int_{-R_1}^{z}dz'e^{-q_n(z-z')}I^1_{n\ell'm'}(z')\pm\int_{z}^{R_1}dz'e^{-q_n(z'-z)}I^1_{n\ell'm'}(z')\right]$$
$$\pm e^{im\tilde{\varphi}_n}e^{-im'\tilde{\varphi}_n}\int_{-R_2}^{-R_1}dz\left[I^{pa}_{n\ell m}(z)\right]^*e^{q_nz}\int_{-R_1}^{R_1}dz'e^{-q_nz'}I^1_{n\ell'm'}(z')$$
$$+e^{im\tilde{\varphi}_n}e^{-im'\tilde{\varphi}_n}\int_{R_1}^{R_2}dz\left[I^{pa}_{n\ell m}(z)\right]^*e^{-q_nz}\int_{-R_1}^{R_1}dz'e^{q_nz'}I^1_{n\ell'm'}(z'); p=2,3$$

(19a)

$$M_{n\ell m \ell' m'}^{\{2\}q\pm} = e^{im\tilde{\varphi}_n} e^{-im'\tilde{\varphi}_n} \int_{-R_1}^{R_1} dz \left[ I_{n\ell m}^{\{2\}b}(z) \right]^* \left[ \int_{-R_1}^{z} dz' e^{-q_n(z-z')} I_{n\ell' m'}^{qb}(z') \pm \int_{z}^{R_1} dz' e^{-q_n(z'-z)} I_{n\ell' m'}^{qb}(z') \right]$$

$$+ e^{im\tilde{\varphi}_n} e^{-im'\tilde{\varphi}_n} \int_{-R_2}^{-R_1} dz \left[ I_{n\ell m}^{\{2\}a}(z) \right]^* \left[ \int_{-R_2}^{z} dz' e^{-q_n(z-z')} I_{n\ell' m'}^{qa}(z') \pm \int_{z}^{-R_1} dz' e^{-q_n(z'-z)} I_{n\ell' m'}^{qa}(z') \right]$$

$$+ e^{im\tilde{\varphi}_n} e^{-im'\tilde{\varphi}_n} \int_{R_1}^{R_2} dz \left[ I_{n\ell m}^{\{2\}a}(z) \right]^* \left[ \int_{R_1}^{z} dz' e^{-q_n(z-z')} I_{n\ell' m'}^{qa}(z') \pm \int_{z}^{R_2} dz' e^{-q_n(z'-z)} I_{n\ell' m'}^{qa}(z') \right]$$

$$\pm e^{im\tilde{\varphi}_n} e^{-im'\tilde{\varphi}_n} \int_{-R_2}^{-R_1} dz \left[ I_{n\ell m}^{\{2\}a}(z) \right]^* e^{q_n z} \int_{R_1}^{R_2} dz' e^{-q_n z'} I_{n\ell' m'}^{qa}(z')$$

$$+ e^{im\tilde{\varphi}_n} e^{-im'\tilde{\varphi}_n} \int_{R_1}^{R_2} dz \left[ I_{n\ell m}^{\{2\}a}(z) \right]^* e^{-q_n z} \int_{-R_2}^{-R_1} dz' e^{q_n z'} I_{n\ell' m'}^{qa}(z')$$

$$\pm e^{im\tilde{\varphi}_n} e^{-im'\tilde{\varphi}_n} \int_{-R_2}^{-R_1} dz \left[ I_{n\ell m}^{\{2\}a}(z) \right]^* e^{q_n z} \int_{-R_1}^{R_1} dz' I_{n\ell' m'}^{qb}(z') e^{-q_n z'}$$

$$+ e^{im\tilde{\varphi}_n} e^{-im'\tilde{\varphi}_n} \int_{R_1}^{R_2} dz \left[ I_{n\ell m}^{\{2\}a}(z) \right]^* e^{-q_n z} \int_{-R_1}^{R_1} dz' I_{n\ell' m'}^{qb}(z') e^{q_n z'}$$

$$+ e^{im\tilde{\varphi}_n} e^{-im'\tilde{\varphi}_n} \int_{-R_1}^{R_1} dz \left[ I_{n\ell m}^{\{2\}b}(z) \right]^* e^{-q_n z} \int_{-R_2}^{-R_1} dz' I_{n\ell' m'}^{qa}(z') e^{q_n z'}$$

$$\pm e^{im\tilde{\varphi}_n} e^{-im'\tilde{\varphi}_n} \int_{-R_1}^{R_1} dz \left[ I_{n\ell m}^{\{2\}b}(z) \right]^* e^{q_n z} \int_{R_1}^{R_2} dz' I_{n\ell' m'}^{qa}(z') e^{-q_n z'}$$

$; p, q = 2, 3$

(20a)

For the case as shown in Fig. 1(b):

$$M_{n\ell m \ell' m'}^{11\pm} = e^{im\tilde{\varphi}_n} e^{-im'\tilde{\varphi}_n} \int_{-R_1}^{R_1} dz \left[ I_{n\ell m}^{1}(z) \right]^* \left[ \int_{-R_1}^{z} dz' e^{-q_n(z-z')} I_{n\ell' m'}^{1}(z') \pm \int_{z}^{R_1} dz' e^{-q_n(z'-z)} I_{n\ell' m'}^{1}(z') \right],$$

(17b)

$$M_{n\ell m \ell' m'}^{1q\pm} = e^{im\tilde{\varphi}_n} e^{-im'\tilde{\varphi}_n} \int_{-R_1}^{R_1} dz \left[ I_{n\ell m}^{1}(z) \right]^* \left[ \int_{-R_1}^{z} dz' e^{-q_n(z-z')} I_{n\ell' m'}^{qb}(z') \pm \int_{z}^{R_1} dz' e^{-q_n(z'-z)} I_{n\ell' m'}^{qb}(z') \right]$$

$$\pm e^{im\tilde{\varphi}_n} e^{-im'\tilde{\varphi}_n} \int_{-R_1}^{R_1} dz \left[ I_{n\ell m}^{1}(z) \right]^* e^{q_n z} \int_{R_1}^{R_2} dz' e^{-q_n z'} I_{n\ell' m'}^{qa}(z'); q = 2, 3$$

(18b)

$$M_{n\ell m \ell' m'}^{p1\pm} = e^{im\tilde{\varphi}_n} e^{-im'\tilde{\varphi}_n} \int_{-R_1}^{R_1} dz \left[ I_{n\ell m}^{pb}(z) \right]^* \left[ \int_{-R_1}^{z} dz' e^{-q_n(z-z')} I_{n\ell' m'}^{1}(z') \pm \int_{z}^{R_1} dz' e^{-q_n(z'-z)} I_{n\ell' m'}^{1}(z') \right]$$

$$+ e^{im\tilde{\varphi}_n} e^{-im'\tilde{\varphi}_n} \int_{R_1}^{R_2} dz \left[ I_{n\ell m}^{pa}(z) \right]^* e^{-q_n z} \int_{-R_1}^{R_1} dz' e^{q_n z'} I_{n\ell' m'}^{1}(z'); p = 2, 3$$

(19b)

$$M_{n\ell m\ell'm'}^{\{^2_3\}q\pm} = e^{im\tilde{\varphi}_n}e^{-im'\tilde{\varphi}_n}\int_{-R_1}^{R_1}dz\left[I_{n\ell m}^{\{^2_3\}b}(z)\right]^*\left[\int_{-R_1}^{z}dz'e^{-q_n(z-z')}I_{n\ell'm'}^{qb}(z') \pm \int_{z}^{R_1}dz'e^{-q_n(z'-z)}I_{n\ell'm'}^{qb}(z')\right]$$

$$+e^{im\tilde{\varphi}_n}e^{-im'\tilde{\varphi}_n}\int_{R_1}^{R_2}dz\left[I_{n\ell m}^{\{^2_3\}a}(z)\right]^*\left[\int_{R_1}^{z}dz'e^{-q_n(z-z')}I_{n\ell'm'}^{qa}(z') \pm \int_{z}^{R_2}dz'e^{-q_n(z'-z)}I_{n\ell'm'}^{qa}(z')\right]$$

$$+e^{im\tilde{\varphi}_n}e^{-im'\tilde{\varphi}_n}\int_{R_1}^{R_2}dz\left[I_{n\ell m}^{\{^2_3\}a}(z)\right]^* e^{-q_n z}\int_{-R_1}^{R_1}dz'I_{n\ell'm'}^{qb}(z')e^{q_n z'}$$

$$\pm e^{im\tilde{\varphi}_n}e^{-im'\tilde{\varphi}_n}\int_{-R_1}^{R_1}dz\left[I_{n\ell m}^{\{^2_3\}b}(z)\right]^* e^{q_n z}\int_{R_1}^{R_2}dz'I_{n\ell'm'}^{qa}(z')e^{-q_n z'}$$

$;p,q=2,3$

(20b)

For the $\delta(z-z')$ term, we have:

For the case as shown in Fig. 1(a):

$$M_{n\ell m\ell'm'}^{110} = e^{-im'\tilde{\varphi}_n}e^{im\tilde{\varphi}_n}\int_{-R_1}^{R_1}dz\left[I_{n\ell m}^{1}(z)\right]^* I_{n\ell'm'}^{1}(z), \tag{21a}$$

$$M_{n\ell m\ell'm'}^{1q0} = e^{-im'\tilde{\varphi}_n}e^{im\tilde{\varphi}_n}\int_{-R_1}^{R_1}dz\left[I_{n\ell m}^{1}(z)\right]^* I_{n\ell'm'}^{qb}(z); q=2,3, \tag{22a}$$

$$M_{n\ell m\ell'm'}^{p10} = e^{-im'\tilde{\varphi}_n}e^{im\tilde{\varphi}_n}\int_{-R_1}^{R_1}dz\left[I_{n\ell m}^{pb}(z)\right]^* I_{n\ell'm'}^{1}(z); p=2,3, \tag{23a}$$

$$M_{n\ell m\ell'm'}^{pq0} = e^{-im'\tilde{\varphi}_n}e^{im\tilde{\varphi}_n}\int_{-R_2}^{-R_1}dz\left[I_{n\ell m}^{pa}(z)\right]^* I_{n\ell'm'}^{qa}(z) + e^{-im'\tilde{\varphi}_n}e^{im\tilde{\varphi}_n}\int_{-R_1}^{R_1}dz\left[I_{n\ell m}^{pb}(z)\right]^* I_{n\ell'm'}^{qb}(z)$$
$$+e^{-im'\tilde{\varphi}_n}e^{im\tilde{\varphi}_n}\int_{R_1}^{R_2}dz\left[I_{n\ell m}^{pa}(z)\right]^* I_{n\ell'm'}^{qa}(z); p,q=2,3$$

(24a)

For the case as shown in Fig. 1(b):

$$M_{n\ell m\ell'm'}^{110} = e^{-im'\tilde{\varphi}_n}e^{im\tilde{\varphi}_n}\int_{-R_1}^{R_1}dz\left[I_{n\ell m}^{1}(z)\right]^* I_{n\ell'm'}^{1}(z), \tag{21b}$$

$$M_{n\ell m\ell'm'}^{1q0} = e^{-im'\tilde{\varphi}_n}e^{im\tilde{\varphi}_n}\int_{-R_1}^{R_1}dz\left[I_{n\ell m}^{1}(z)\right]^* I_{n\ell'm'}^{qb}(z); q=2,3, \tag{22b}$$

$$M_{n\ell m\ell'm'}^{p10} = e^{-im'\tilde{\varphi}_n}e^{im\tilde{\varphi}_n}\int_{-R_1}^{R_1}dz\left[I_{n\ell m}^{pb}(z)\right]^* I_{n\ell'm'}^{1}(z); p=2,3, \tag{23b}$$

$$M_{n\ell m\ell'm'}^{pq0} = e^{-im'\tilde{\varphi}_n}e^{im\tilde{\varphi}_n}\int_{-R_1}^{R_1}dz\left[I_{n\ell m}^{pb}(z)\right]^* I_{n\ell'm'}^{qb}(z) + e^{-im'\tilde{\varphi}_n}e^{im\tilde{\varphi}_n}\int_{R_1}^{R_2}dz\left[I_{n\ell m}^{pa}(z)\right]^* I_{n\ell'm'}^{qa}(z)$$
$;p,q=2,3$

(24b)

Finally, the coefficients $\boldsymbol{\alpha}_{\ell m}, \boldsymbol{\beta}_{\ell m}$, and $\boldsymbol{\gamma}_{\ell m}$ in Eq. (3) can be solved by using an iterative solver such as the quasi-minimum residue (QMR) method. Note that the theoretical formulation of light scattering from both structures as shown in Fig. 1(a) and Fig. 1(b) are very similar. The only difference is that we do not consider the contribution from the interval from $-R_2$ to $-R_1$ for variable $z$. Hence the explicit forms in the above equations for Fig. 1(b) are more complicated.

For a general distribution of identical nanoparticles, Eq. (1) should be replaced by

$$\mathbf{E}_i(\mathbf{r}_i) = \mathbf{E}_0(\mathbf{r}_i) + k_0^2 \sum_j [(\varepsilon_1 - \varepsilon_a) \int_{\Omega_{1j}} \mathbf{G}(\mathbf{r},\mathbf{r}')\cdot\mathbf{E}_j(\mathbf{r}')d^3\mathbf{r}' + (\varepsilon_2 - \varepsilon_a) \int_{\Omega_{2j}} \mathbf{G}(\mathbf{r},\mathbf{r}')\cdot\mathbf{E}_j(\mathbf{r}')d^3\mathbf{r}'], \quad (25)$$

where $\mathbf{E}_i(\mathbf{r}_i)$ denotes the electric field in a local volume surrounding particle $i$, and $\mathbf{r}_i = \mathbf{r} - \mathbf{R}_i$ with $\mathbf{R}_i$ denoting the center position of particle $i$. For a periodic array of nanoparticles, according to the Bloch theorem we have $\mathbf{E}_i(\mathbf{r}_i) = e^{i\mathbf{k}_0\cdot\mathbf{r}}u(\mathbf{r}_i)$, where $u(\mathbf{r}_i)$ is a periodic function (i.e. same for all sites). For a random distribution of nanoparticles, we may approximate $\mathbf{E}_j(\mathbf{r}_j)$ by $f\, e^{i\mathbf{k}_0\cdot\mathbf{r}}u(\mathbf{r}_i)$ for sites $j$ other than $i$, where $f$ denotes a similarity factor.[26,27] For a completely random and uniform distribution, every nanoparticle "sees" the same environment. Thus, the site function $u(\mathbf{r}_i)$ becomes the same for all sites on average (i.e. $f = 1$) For a random but non-uniform distribution, the introduction of a similarity factor $f$ less than 1 simulates the average behavior of the dissimilarity between site functions. With this approximation, the site average of Eq. (25) is then replaced by a scattering equation similar to Eq. (1) [26,27] with $\mathbf{G}(\mathbf{r},\mathbf{r}')$ replaced by the average Green's function

$$\bar{\mathbf{G}}(\mathbf{r},\mathbf{r}') = \frac{1}{(2\pi)^2} \int d\mathbf{k}_n e^{i\mathbf{k}_n\cdot(\boldsymbol{\rho}-\boldsymbol{\rho}')} \mathbf{g}_n(z,z') S(\mathbf{K}_n), \quad (26)$$

where $\mathbf{K}_n = \mathbf{k}_n - \mathbf{k}_0$ and $S(\mathbf{K}_n)$ denotes the structure factor defined as [26,27]:

$$S(\mathbf{K}_n) = 1 + \frac{1}{N} f \sum_{j\neq 1} e^{-i\mathbf{K}_n\cdot(\mathbf{R}_j-\mathbf{R}_1)} e^{-(\mathbf{R}_j-\mathbf{R}_i)^2/2\lambda_c^2} \approx 1 + \frac{2\pi f \lambda_c^2}{A_{cell}} - \frac{2\pi f}{A_{cell}} \int_0^{R_u} J_0(K_n R) R dR. \quad (27)$$

where $R_u$ is the average cell radius, and $A_{cell} = \pi R_u^2$ is the average cell area.

## III. Comparison with the core-shell Mie scattering theory

To check the accuracy of our model calculation, we compare our calculated results based on GF method with the Mie scattering theory for an isolated core-shell nanosphere [30]. The geometry is shown in Fig. 1(a) and we set $\varepsilon_{sub}=1$ for the substrate. We first consider a core made of Au and the shell made of a dielectric with $\varepsilon=2$. The background material is air. The dash-dotted line, dotted line and dashed line denote results obtained by the newly Green's function method with various cutoff values of angular momentum quantum number $\ell$ ($\ell_c$), number of $k_n$ mesh ($N_k$), and number of z mesh ($N_z$): $(\ell_c, N_k, N_z) = (5,51,50), (5,51,100), (5,101,100)$, respectively. The inner radius ($R_1$) and outer radius ($R_2$) considered are 20nm and 40nm, respectively. Fig. 2(a) shows the calculate field strength at three different positions. These three different positions are located at $r = R_1 - \delta$ (A point), $r = R_1 + \delta$ (B point) and $r = R_2 - \delta$ (C point), where $\delta$ is an infinitesimal number. We find the rad dashed line denotes the convergent results by our GF method and find good agreement with the core-shell Mie theory. Next, we consider a Au sphere enclosed by a Ag shell with the same inner and outer radius as before. All other parameters are the same as before. The results are shown in Fig. 2(b), and again good agreement with the Mie theory is obtained. This indicates that our GF method has good enough accuracy for calculating optical scattering from core-shell nanoparticles, and the method is expected to work well in the presence of a substrate. Since the theoretical formulation for a partially-covered core-shell nanoparticle [Fig. 1(b)] is very similar to an isolated core-shell nanoparticle [Fig. 1(a)], we expect the results also work well for partially-covered core-shell nanoparticle on a substrate.

## IV. Application to spectroscopic ellipsometry sensing of DNA molecules by using Au nanoparticles

First, we briefly introduce the experimental process as considered in [28]. The experiment was done by using a variable-angle spectroscopic ellipsometer (VASE, J. A. Woollam Co.) system with an adjustable retarder to conduct the SE measurements with a setup as shown in Fig. 3. Surface modification on clean glass substrates has been done with 10% 3-Aminopropyltrimethoxysilane (APTES) (v/v) solution in ethanol. Gold nanoparticles with average diameter of 13nm were prepared by reduction of chloroauric acid (HAuCl4) solution. The Au nanoparticles solution was poured on the APTES modified glass surface for one hour in a humid environment then rinsed with ultrapure water followed by drying with nitrogen gas. The DNA sequence containing 16 bases single strand oligonucleotides 5´ C T A C C T T T T T T T T C T G 3´ (Thiol (SH) group modified) and 5´ C A G A A A A A A A G G T A

G 3 ´ were used for hybridization. For the DNA hybridization experiments, the DNA molecules were diluted in 5x saline-sodium citrate buffer (SSC, pH= 7.0). Briefly, 1µM of probe SH-DNA molecules was injected onto gold nanoparticle sample surface integrated with micro-fluidic cell. The solution is kept for 3 hours to achieve sufficient coverage of DNA molecules on surface of gold nanoparticle layer. The sample is further rinsed with SSC buffer to remove unbound probe DNAs. Finally, 1 µM of target DNA solution was injected into the micro-fluidic cell and kept for another 5 hours to undergo DNA hybridization. Rinsing process is repeated after the experiment to remove the non-hybridized target DNAs. Figure 4 shows the AFM picture of the hybridized DNA sequence attached on Au nanoparticles.

Next we compare our calculated results with experimental data for a random distribution of Au nanospheres placed on the flat surface along the long edge of a BK7 dove prism. An SEM picture of the sample is shown in Fig. 4. As a first approximation, we model the distribution of nanoparticles by a periodic array. We choose the cutoff of angular momentum quantum number $\ell$ ($\ell_c$), the plane wave number of $k_n \left( k_{nx} \times k_{ny} \right)$ ($N_{k_{nx}} \times N_{k_{ny}}$) and the number of z mesh ($N_z$) to be ($N_{k_{nx}}, N_{k_{ny}}, \ell_c, N_z$) = (41,41,3,50), which has been tested to give convergent results for small nanoparticles as considered here. The best fit to experimental data was obtained by adopting a pitch ($p$) of 30nm for a square unit cell. The model system, as depicted in Fig. 1, consists of three regions: (i) the ambient, which is BK7 prism (since light is assumed to enter the system from BK7) plus a Cauchy layer, describing the bottom of the microfluidic glass chamber plus the index-matching oil) whose thickness is 28.36nm (Note that the dielectric constant of the Cauchy layer is very close to that of BK7; thus, it is a good approximation to consider it as part of the ambient), (ii) a grating layer which contains Au nanoparticles plus DNA shell embedded in water (the host medium), and (iii) the substrate (corresponding to water in the experimental setup). We also note that the *z* is pointing downward in Fig. 1, which is reversed from the plot for experimental setup as shown in Fig. 3. The measured ellipsometric spectra for Ψ and Δ of clustering Au nanoparticles inside the microfluidic setup on a dove prism are shown in Fig. 5(a) (black curves) and the best-fit theoretical results obtained by our GF method are shown in Fig. 5(b) (black curves). We find reasonable agreement between theory and experiment.

Furthermore, we compare our calculated results with experimental data after DNA molecules were captured on the surface of Au nanospheres. We model the DNA coverage by an effective dielectric thin shell which encloses the Au nanosphere as shown in Fig. 1(b) rather than Fig. 1(a) since the mass of Au nanoparticles are attached to the glass before the coverage of DNA molecules. Then, we use the GF method described above to calculate the light scattering from nanoparticles on the substrate partially covered by DNA shells. In this case, the parameters are the same as discussed above. We consider the effective thickness $\delta$ of the thin shell of DNA as adjustable parameter and fit the experimental data for ellipsometric spectra. Fig. 5(a) shows the measured ellipsometric spectra ($\Psi$ and $\Delta$) obtained by VASE for a random distribution of closely-spaced nanospheres on a BK7 prism without (black solid line) and with (red dash-dotted line) the DNA coverage. We observed a small but detectable red shift of the plasmonic peak in both $\Psi$ and $\Delta$ spectra when DNA molecules are attached. We find that the shift in $\Delta$ spectrum is much more pronounced than the $\Psi$ spectrum, indicating a better sensitivity for sensing by using the $\Delta$ signal. Furthermore, we show in Fig. 5(b) the theoretical results with (red dash-dotted line) and without (black solid line) the DNA coverage. The best-fit result (red dash-dotted line) is obtained by using a shell thickness $\delta$ =0.2nm. The dielectric constant of $\varepsilon$ =7.3 is used for the DNA shell which is close to the previous measured dielectric constant for double-strand DNA as a uniform medium [31]. When the nanoparticle is fully wrapped around by DNA molecules, the thickness should be thicker than 0.2nm, since the width of a DNA chain is around 2 to 3nm [32]. Thus, our best fit result for the effective thickness of 0.2nm indicates that the fill factor of DNA molecules in the shell layer is around 0.1. Alternatively, we may choose an effective layer thickness of 2.5 nm and adjust the effective dielectric constant of the DNA filled layer to do the fit. The best fit is obtained with an effective dielectric constant of the loosely filled DNA layer set to be 1.96 as indicated by the blue dashed curve in Fig. 5(b).

To check the reliability of the current GF method, we also performed the finite-element calculation by using the COMSOL package and the results are shown in Fig. 5(c). The geometry and material parameters used are the same as in Fig. 5(b). The results obtained by these two methods agree quite well. The slight difference in spectra near the plasmonic resonance between Figs. 5(b) and 5(c) can be attributed to the difference in numerical accuracy of the two methods. We found that our current GF method is more efficient than the COMSOL package by about one order of magnitude, since the finite-element method requires very high mesh density to simulate the strongly localized electromagnetic fields near the plasmonic resonance. For example, it took about 3 minutes in CPU to generate the spectrum depicted by black-line in Fig. 5 (b) with the current GF method, while it took about 3.3 hours to obtain the corresponding spectrum in Fig. 5(c) by using COMSOL with a single Intel 1.7GHz i5 processor.

For generating the colored lines (with DNA coverage) in Fig. 5 (b) and (c), it took about 13 minutes with the GF method and about 4 hours with the COMSOL package.

Moreover, the random distribution of nanoparticles can be easily taken into account in our GF method (without increasing computational cost). We have also tested the FDTD method by using a split-field approach and found that it is much more time consuming than COMSOL and GF method, and it does not get as accurate ellipsometry spectra, since it is difficult to incorporate the frequency-dependent dielectric function in FDTD method. Furthermore, when the angle of incidence is as large as 72.8 degrees considered here, it is more difficult to achieve convergent result with the split-field FDTD simulation since there exists a constraint between the angle of incidence and the time-step. In order to speed up our split-field FDTD method, we performed the calculation with a GPU processor, a massive parallel framework under the CUDA architecture, which is widely recognized to boost the computation by more than one order of magnitude. To achieve similar results, the GPU computation time needed is about 1.5 hours, which translates to more than 15 hours if a single CPU processor was used.

The close agreement between our GF method and COMSOL package indicates that our current GF approach for treating core-shell nanoparticles is reliable. However, since the distribution of nanoparticles as shown in Fig. 4 is not periodic, it is desirable to consider the effect due to random distribution of Au nanoparticle by introducing the structure factor $S(\mathbf{K}_n)$ as described in Eq. (27). In our modeling, the cutoff of angular momentum quantum number $\ell$ ($\ell_c$), the number of $k_n$ meshes ($N_k$), and the number of z meshes ($N_z$) used are ($\ell_c, N_k, N_z$) = (3,51,50). The similarity factor $f$ used in the fitting is 0.7, and the coherent length of light $\lambda_c$ is 5000nm. (Note that the results are insensitive to $\lambda_c$ as long as it is longer than 3000nm) The results obtained are shown in Fig. 5(d). Note that the average pitch ($p=\sqrt{A_{cell}}$) used is 30 nm, which is the same as the one in Fig. 5(b).

As seen in Fig. 5(d), including the effect of random distribution gives closer agreement with the experimental results shown in Fig. 5(a) in that the sharp shoulder near 570nm in the $\Delta$ spectrum is smoothed out, which resembles the experimental data better. As for computational time, it took only 30 (210) seconds in CPU time to generate the spectrum for the random distribution of Au nanoparticles without (with) DNA coverage as shown by the black (colored) line in Fig. 5(d) by using a single Intel 1.7GHz i5 processor. So, our GF method is especially effective for treating random distribution of core-shell nanoparticles on a substrate.

The remaining discrepancy between theory and experiment may be attributed to the size non-uniformity and the clustering effect, which are not yet included in the present simulation.

## V. Conclusion

In this paper, we have constructed the scattering theory for partially-covered core-shell particles on a substrate. We have applied our theory to model the DNA coverage on metallic spheres. We illustrate how to use the model calculation based on the half-space Green's function approach to interpret the ellipsometric spectra of randomly distributed coupled Au nanospheres placed on a BK7 dove prism. We found fairly good agreement between the model calculation and experiment. Furthermore, we have used spectroscopic ellisometry to detect the coverage of Au nanoparticles by DNA molecules. We found a small but detectable spectroscopic shift in both the Ψ and Δ spectra with more significant change in Δ spectra in both experimental and theoretical results. We model the DNA coverage by a thin dielectric shell which encloses the Au nanoparticle and then use a Green's function method to investigate light scattering from coupled Au nanoparticles on a substrate, partially covered by shells of biomolecules. By comparing the model calculation with experiment, we found that the DNA coverage causes a small red shift of the plasmonic peak, and the effective thickness of DNA coverage is around 0.2nm, when the dielectric constant of the double-strand DNA is adopted [31]. In an alternative study, we keep the shell thickness at 2.5 nm (close to the width of a DNA chain [32]), and we obtain an effective dielectric constant of 1.96 for the layer partially filled with DNA molecules. We have compared our simulation results with the ones calculated by using the COMSOL package and found close agreement, while our method is about one order of magnitude more efficient than the COMSOL package. Furthermore, we also considered the effect due to random distribution of nanoparticles and found improved agreement between the model calculation and experiment. Hence, we have demonstrated that the spectroscopic ellipsometric measurements coupled with theoretical analysis via an efficient computation method can be an effective tool for detecting DNA molecules attached on Au nanoparticles, thus achieving label-free, non-destructive, and high-sensitivity biosensing with nanometer resolution.

**Acknowledgements**

This work was supported in part by the Nanoproject of Academia Sinica and National Science Council of Taiwan under Contract No. NSC 101-2112-M-001-024-MY3.

**Figure captions**

**Figure** 1. Schematic diagram of light scattering from (a) a core-shell nanoparticle and (b) a partially-covered core-shell nanoparticle on a substrate. $\varepsilon_a$, $\varepsilon_g$, $\varepsilon_1$, $\varepsilon_2$, and $\varepsilon_{sub}$ denote the dielectric constants of the ambient, host material of grating layer, core region ($\Omega_1$), shell region ($\Omega_2$), and substrate, respectively.

**Figure** 2. The calculate field strength, $|\mathbf{E}|$ at position A, B and C of the isolated core-shell nanosphere as a function of photon energy for light scattering from an isolated Au nanosphere obtained by both Mie scattering theory (black solid line) and the current Green's-function approach with three different sets of cutoff parameters: $(\ell_c, N_k, N_z) = (5, 51, 50)$ (blue dash-dotted), $(5, 51, 100)$ (green dotted), and $(5, 101, 100)$ (red dash-dotted). The dielectric constant of the shell used is (a) $\varepsilon = 2$ and (b) a frequency-dependent $\varepsilon_{Ag}(\omega)$ for silver.

**Figure** 3. Optical setup of surface plasmon resonance ellipsometry used in the present study with the illustration of polarization of incident and reflected light within the prism (treated as ambient).

**Figure** 4. AFM image of DNA molecules attached on Au nanoparticles with diameter 13 nm.

**Figure** 5. (a) Experimental ellisometry spectra (Ψ and Δ) obtained by VASE for a random distribution of Au nanoparticles placed on a BK7 prism without (black line) and with (red dash-dotted line) the DNA molecules attached. (b) Calculated ellisometry spectra (Ψ and Δ) obtained by the current GF method for a periodic array of nanopartciles without (black line) and with DNA coverage (red and blue lines). (c) Calculated ellisometry spectra (Ψ and Δ) of a periodic array of Au nanoparticles obtained by using the COMSOL package without (black line) and with DNA coverage (red and blue lines). The same two different sets of fitting parameters as in (b) were used. (d) Calculated ellisometry spectra ($\Psi$ and $\Delta$) for a random distribution of Au nanoparticles obtained by the current GF method without (black line) and with DNA coverage (red and blue lines). In (b)-(d) two different sets of fitting parameters are chosen. For the red dash-dotted line, the dielectric constant of DNA shell used is 7.3 [31] and the shell thickness is 0.2nm. For the blue dashed line, the thickness of DNA shell used is 2.5nm [32] and the effective dielectric constant of the DNA shell is 1.96.

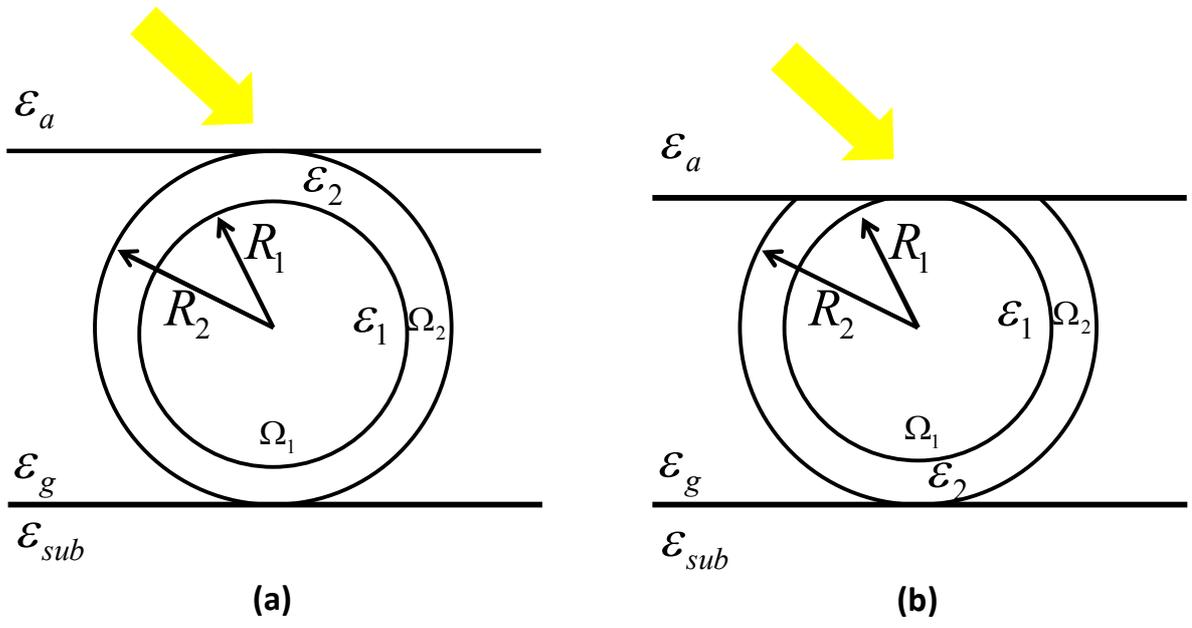

Fig. 1

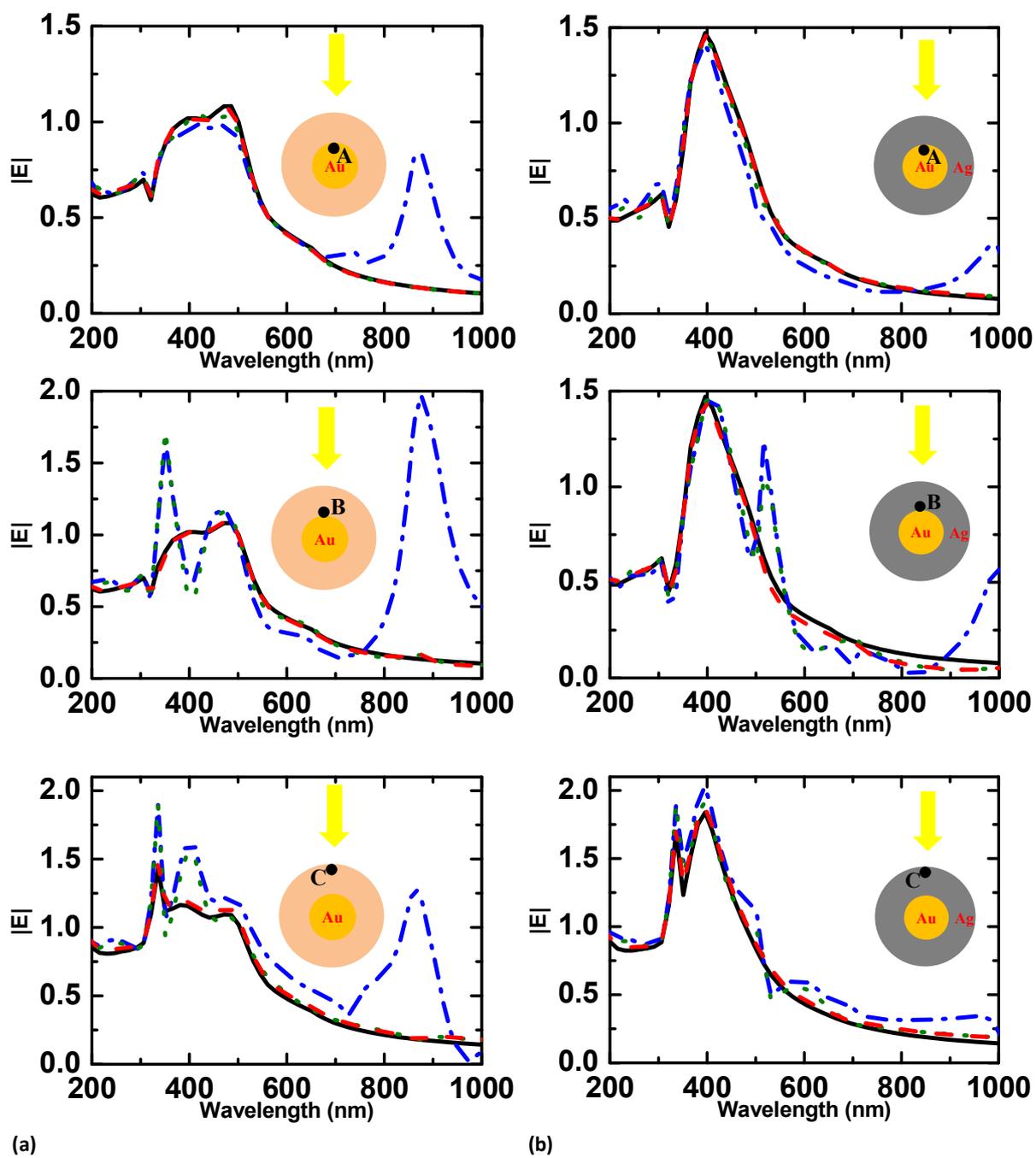

**Fig. 2**

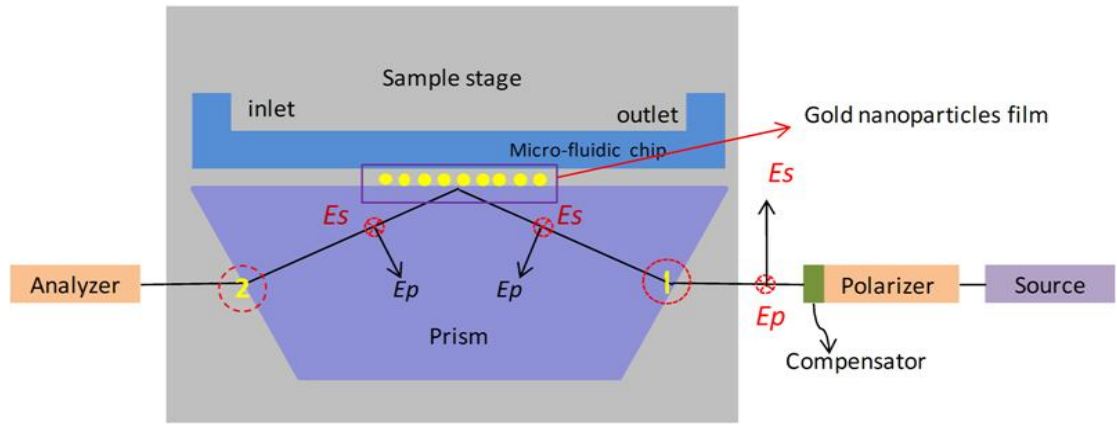

**Fig. 3**

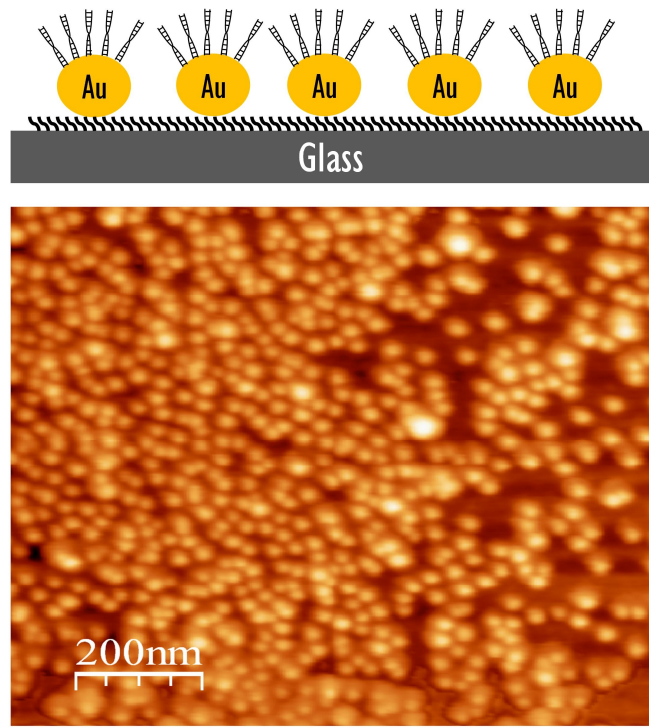

**Fig. 4**

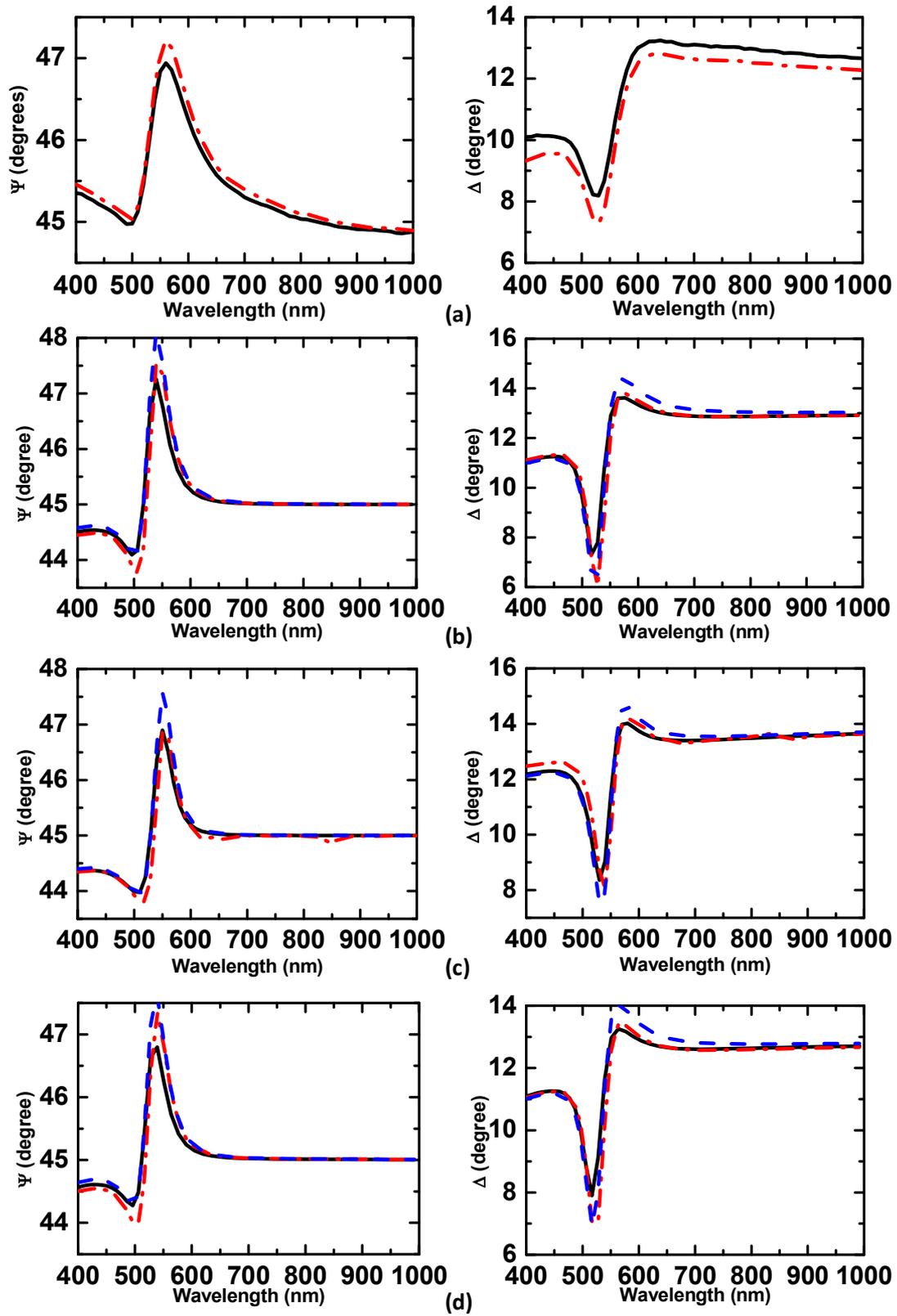

Fig. 5